\documentclass[a4,useAMS,usenatbib,usegraphicx]{mn2e}
\def\aapr{\ref@jnl{A\&A~Rev.}}		

\usepackage{amsmath}
\include{epsf}
\usepackage{float}
\usepackage{color}

\usepackage{epsf}
\epsfclipon

\title[Gas expulsion timescales]{Infant mortality in the hierarchical merging scenario: Dependency on gas expulsion timescales}
\author[R.Smith et al]{R.Smith$^{1}$\thanks{E-mail:rsmith@astro-udec.cl}, S. Goodwin${^2}$, M. Fellhauer${^1}$, P. Assmann${^1}$\\
$^{1}$Departamento de Astronomia, Universidad de Concepcion, Casilla 160-C, Concepcion, Chile\\
\noindent
$^{2}$Department of Physics and Astronomy, University of Sheffield, Hicks Building, Hounsfield Road, Sheffield, S3 7RH, UK}
\begin{document}

\date{Accepted to MNRAS, October 1st 2012}

\pagerange{\pageref{firstpage}--\pageref{lastpage}} \pubyear{2011}

\maketitle

\label{firstpage}

\begin{abstract}
We examine the effects of gas expulsion on initially sub-structured
and out-of-equilibrium star clusters.  We perform $N$-body simulations
of the evolution of star clusters in a static background potential
before adjusting that potential to model gas expulsion. We investigate
the impact of varying the rate at which the gas is removed, and the
instant at which gas removal begins.

Reducing the rate at which the gas is expelled results in an increase
in cluster survival. Quantitatively, this dependency is approximately
in agreement with previous studies, despite their use of smooth, and
virialised initial stellar distributions.

However, the instant at which gas expulsion occurs is found to have a
strong effect on cluster response to gas removal. We find if gas
expulsion occurs prior to one crossing time, cluster response is
poorly described by {\it{any}} global parameters. Furthermore in real
clusters the instant of gas expulsion is poorly constrained.
Therefore our results emphasis the highly stochastic and variable
response of star clusters to gas expulsion.
\end{abstract}


\section{Introduction}
Many young stars are found in clusters.  The fraction which are found
in clusters depends on ones definition of a cluster (see Lada \& Lada
2003; Bressert et al. 2010), and how exactly a cluster is defined
(bound vs. unbound, see e.g. \citealp{Gieles2011}).  However
one looks at it though, star `clusters' are clearly an important mode of
star formation, if not the dominant mode.  

Most stars appear to form in groups of tens to thousands of 
members embedded in molecular clouds.  The initial distribution of
stars follows the complex clumpy and filamentary structure of the
underlying gas (see e.g. \citealp{Kirk2007}; \citealp{Gutermuth2009}; 
\citealp{Peretto2009}; \citealp{Bressert2010}; 
\citealp{Difrancesco2010}; \citealp{Maury2011}). Simulations of initially smooth clouds of molecular gas, that are seeded with supersonic turbulence, fragment into filamentary structures (\citealp{Bonnell2003}; \citealp{Bate2004}; \citealp{Bate2009}; \citealp{Girichidis2011}). These structures may collapse to form stars if they become self-gravitating.
It has been argued that clusters form as entities within this 
complex distribution (e.g. Marks \& Kroupa 2011), or 
that they can form afterwards by
mergers of stars and stellar groups from this complex distribution
(\citealp{Allison2009b}; \citealp{Allison2010}). Clearly the initial dynamical state of
the young stars will play a crucial role in forming bound clusters or
unbound associations (\citealp{Allison2009b}; \citealp{Gieles2011}).  

In this paper we will take the view that clusters form within the 
hierarchical merging scenario (\citealp{Allison2009b}; \citealp{Allison2010}) and, together
with gas not used in star formation, they form bound objects.
Observations show very few young clusters associated with natal gas
older than $\sim 5$~Myr (\citealp{Lada2003}; \citealp{Lada2010}).  It is assumed
that feedback from massive stars removes residual gas in a gas
expulsion phase.  This gas expulsion will significantly alter the
potential felt by the stars and can result in the destruction of the
cluster (e.g. \citealp{Tutukov1978}; \citealp{Hills1980}; \citealp{Goodwin2006};
\citealp{Baumgardt2007}; \citealp{Goodwin2009}).  This is often cited as the
cause of `infant mortality': the apparently high destruction rate of
young clusters\footnote{Although it should be noted that the
  importance of infant mortality depends on one's definition of a
  cluster.} (e.g. \citealp{Lada2003}).

Previous work on gas expulsion has tended to concentrate on clusters
in which the gas and stars are in virial and dynamical equilibrium
(but see \citealp{Verschueren1989}; \citealp{Goodwin2009}).  
The assumption of virial and dynamical equilibrium for 
young clusters can be used 
to infer the initial cluster population and properties
from present-day populations (e.g. \citealp{Parmentier2007}; 
\citealp{Baumgardt2008}; \citealp{Parmentier2008}).  However, the
assumption of equilibrium initial conditions 
means that there is a one-to-one correlation between
survivability and the initial conditions.

This paper is part
of a series in which we examine gas expulsion in the context of the
hierarchical merging scenario in which the initial distributions of
the stars and gas are not in dynamical equilibrium. Specifically they
have an initial distribution which will cause them to collapse and
form a bound cluster within the gas potential (\citealp{Allison2009b}; \citealp{Allison2010}; Smith et al. 2011a,b).

In this paper we present a simple numerical experiment examining the
evolution of non-equilibrium small-$N$ clusters within a smooth
background potential which models the gas.  Whilst this is clearly not
realistic (at birth stars follow a gas distribution which is extremely
complex, see above) we wish to examine the stochasticity which complex
stellar distributions introduce.  As we shall see, statistically
identical non-equilibrium clusters can evolve in {\em very} different 
ways even in our very simple numerical experiments\footnote{And it 
is difficult to see how the addition of
more realistic physics can make the problem less stochastic 
rather than more.} and this stochasticity removes the simple
correlation between initial conditions and survivability that is so
often assumed.

In this paper we examine the $N$-body  evolution of highly non-equilibrium
star clusters in background potentials designed to model the residual
gas in young star  clusters.  Here we adjust the background potential
to model the effects of gas expulsion on different timescales,
starting at different times.  We describe our initial conditions in
Section~2, our results in Section~3, and then discuss the potential
consequences in Section~4 before drawing our conclusions in Section~5.

\section{Initial conditions}

In this paper we present a very simple numerical experiment:
a clumpy distribution of equal-mass stars moving in a smooth (Plummer)
background potential which is then removed to simulate gas expulsion.
This experiment is {\em not} meant to accurately reflect reality, rather it
is meant to concentrate on a new important variable: complex initial
stellar distributions which can introduce (very significant)
stochasticity to the problem of gas expulsion.  As we will discuss in
Section~4 we feel it emphasises some crucial physical parameters
that would be important in any more sophisticated simulations 
and, most importantly, in reality.

We perform our $N$-body simulations using the {\sc{Nbody6}} code
(\citealp{Aarseth2003}). In our simulations there are  two separate
mass distributions which we wish to model: the stars and the
background gas potential.

\subsection{Initial distributions}

In all cases we model the stellar distribution as $N=1000$ particles
with equal masses of $0.5 M_\odot$ resulting in a total stellar mass
of $500 M_\odot$. We choose equal-mass particles in order to avoid
complex two-body interactions and mass segregation. \cite{Allison2009b}
and \cite{Allison2010} showed that both of these effects can be
extremely important in the violent collapse of cool, clumpy regions.
However we wish to avoid complicating our simulations with these
effects as we are interested in the effects of gas expulsion.

We distribute the stars within a radius of 1.5~pc with a fractal
distribution. We choose the fractal dimension $D$ to be $D=1.6$. The fractal
is constructed using the box fractal method described in detail in
Goodwin \& Whitworth 2004 (see also \citealp{Allison2010}).  A fractal
with $D=1.6$ corresponds to a highly clumpy initial distribution (see
upper-left panel of Figure \ref{reprun}). It should be noted that clumpy
fractal clusters  can vary considerably in appearance depending on the
random realisation used and their subsequent evolution can be highly
stochastic (see \citealp{Allison2010}; Smith et al. 2011a). When
investigating their response to varying gas expulsion time-scales, we
therefore conduct a minimum of 5 realisations of any cluster.

The gas within the star-forming region is modelled as a background
Plummer potential with a mass $M_{\rm{g}}=3450$~$M_\odot$ and
scale-radius $r_{\rm{g}}=1.0$~pc. The crossing-time of the region is
$t_{\rm{cr}}\sim 1.3~$Myr.

We emphasise that the gas potential does not follow the initial
stellar distribution, nor does it react to changes in the stellar
distribution as it evolves (it is not live).  These are obviously
extreme simplifications but, as we will discuss later, we feel that we
capture the essence of the basic physics using such a simple model.

We set the initial velocity dispersion of the stars relative to the
total potential (gas \& stars) with initial stellar virial ratios,
$Q_{\rm{i}}$, of 0.2 (sub-virial), or 0.5 (virialised).  Note that a
clumpy stellar distribution with $Q_{\rm{i}} = 0.5$ is {\em not}
initially in dynamical equilibrium. Thus the cluster will
evolve significantly due to violent relaxation.

\subsection{Star formation efficiency}

The choice of the initial mass ($M_{\rm{g}}$) and Plummer radius
($r_{\rm{g}}$) of the (Plummer) gas potential 
sets the {\em true} star formation
efficiency (SFE), i.e. initial fraction of gas that
has been turned into stars.  For a gas mass of
$M_{\rm{g}}=3450$~$M_\odot$, and scale radius $r_{\rm{g}}=1.0$~pc, within 1.5~pc the enclosed gas mass is 2000~$M_\odot$. The stars are distributed out to 1.5~pc, with a total stellar mass of $M_{\rm{s}}=500
~M_\odot$. Thus within a radius of 1.5~pc the true SFE is 20 per cent.

We note that this is below the critical SFE for the survival of a
bound core from an initially virialised star-gas distribution with
instantaneous gas expulsion ($\sim 30$ per cent,
\citealp{Goodwin2006}; \citealp{Baumgardt2007}), and at the lower
bound for the survival of a bound core for adiabatic gas expulsion
(\citealp{Baumgardt2007}).  

\subsection{Gas expulsion}
Although young star clusters form embedded within the molecular
gas from which they formed, few star clusters over $\sim 5$~Myr old
remain associated with their gas (\citealp{Lada2003}; \citealp{Proszkow2009}). This is
likely as a result of a number of mechanisms including radiative
feedback from massive stars, stellar winds from young stars, and
eventually the first supernova(e). The time at which gas removal
begins to occur, and the duration of the gas removal process is
uncertain, and dependent on the particular gas removal mechanism in
operation.

One of the main aims of our simulations is to examine the decoupling
of the gas and stars {\em before} gas expulsion.  If the gas and/or
stars are not in equilibrium either as a whole, or with each other,
then the relative distributions of the gas and stars will change.  As shown in our previous papers (Smith et al. 2011a,b) if the stars can
collapse relative to the gas distribution then the effect of gas
expulsion is less important (see also \citealp{Goodwin2009} and references therein).

To simulate gas expulsion we parameterise the time evolution of the
mass of the gas potential as:

\begin{equation}
\label{gasmasslosseqn}
M_{\rm{g}}(t)=\frac{M_{\rm{g}}(0)}{1+\dot{M}(t-t_{\rm{exp}})}
\end{equation}
\noindent
where $t_{\rm{exp}}$ is time at which gas expulsion begins, and
$\dot{M}$ controls the rate at which gas is removed. 

We choose $\dot{M}$ such that the gas mass $M_{\rm{g}}(t)$ is reduced
to $1~\%$ of its initial value $M_{\rm{g}}(0)$ over a specified
duration time $t_{\rm{dur}}$. For example, if $t_{\rm{exp}}=1~$Myr and
$t_{\rm{dur}}=5$~Myr, then gas expulsion begins 1 Myr after the
beginning of the simulation, and 99$\%$ the gas will be gone 5~Myr
later.

\begin{figure}
  \centering \epsfxsize=8.5cm \epsfysize=7cm
  \epsffile{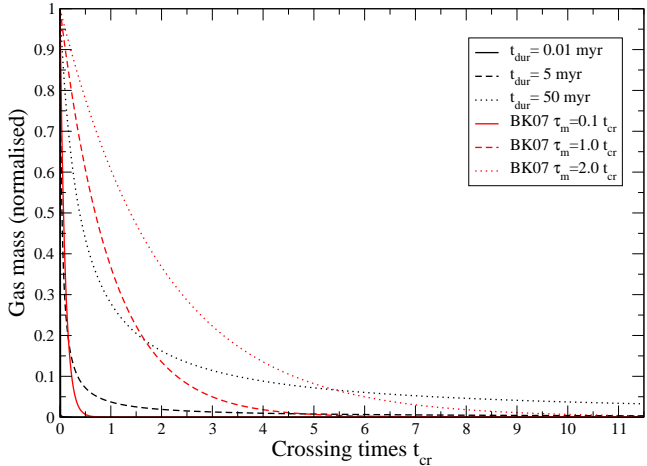}
  \caption{Plot of evolution of gas mass with time using Equation
    \ref{gasmasslosseqn} (black curves), assuming $t_{\rm{cr}}$=1.3~Myr. $t_{\rm{dur}}$ is the length
    of time required for the gas mass to be reduced to 99$\%$ of its
    initial value. When $t_{\rm{dur}}$ is short, gas loss occurs over
    a short duration. Our slowest gas mass-loss is when
    $t_{\rm{dur}}$=50 Myr. For comparison, we include curves of the
    evolution of gas mass from \citet{Baumgardt2007} (red
    curves). $\tau_{\rm{m}}$ is the parameter from
    \citet{Baumgardt2007} controlling the rate of gas loss. We see
    that $t_{\rm{dur}}$=50 Myr results in similar gas
    mass-loss to when $\tau_{\rm{m}}$=1-2 crossing-times.}
\label{gasmasslossfig}
\end{figure}

We test the response of a cluster to four values when gas expulsion
begins ($t_{\rm{exp}}$: 0.1, 1, 5 \& 9 Myr), and four values for how
fast gas expulsion occurs ($t_{\rm{dur}}$: 0.01, 1.0, 5.0 \& 50.0
Myr).  These timescales of gas expulsion cover the reasonable
range of possible gas expulsion timescales from instantaneous to
extremely slow (adiabatic) compared to the crossing time.  It is
unclear what the typical gas expulsion timescales are, and they
may depend on cluster mass (\citealp{Kroupa2002}). For example, the relatively late supernovae of a low-mass O-star might result in a delayed start to gas expulsion.

In Figure \ref{gasmasslossfig}, we illustrate the reduction in gas mass
for gas expulsion that begins effectively immediately ($t_{\rm{exp}}$:
0.1 Myr), and occurs over a duration $t_{\rm{dur}}$: 0.01, 1.0, 5.0 \&
50.0 Myr (black curves). We note that \cite{Baumgardt2007} use an
exponential form for the evolution of the gas mass that differs from
ours in detail, but not in essence. 

\begin{figure*}
\begin{center}$
\begin{array}{cc}
\includegraphics[height=7.2cm,width=9cm]{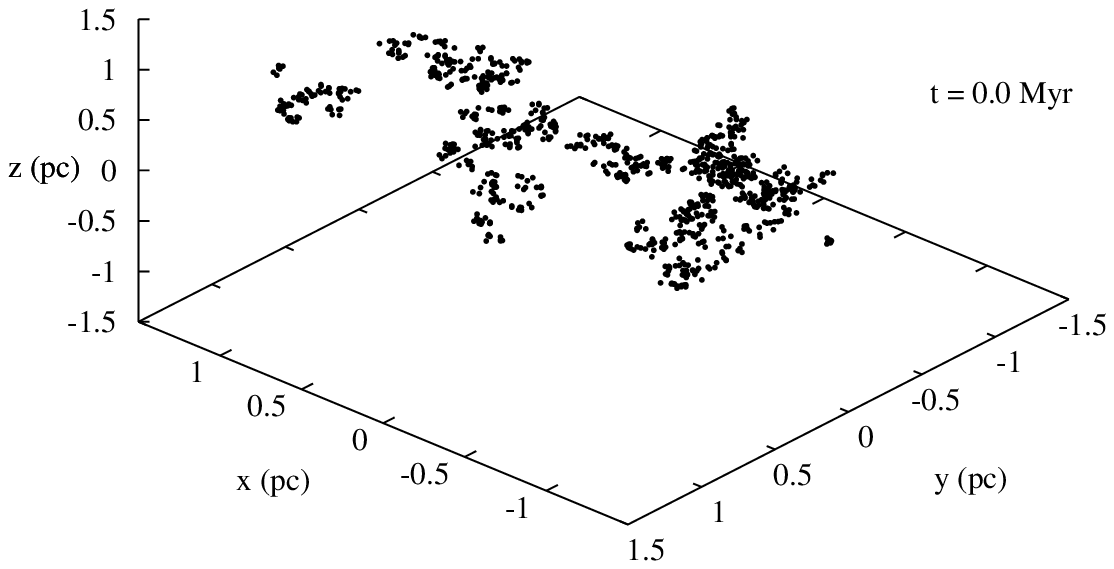} &
\includegraphics[height=7.2cm,width=9cm]{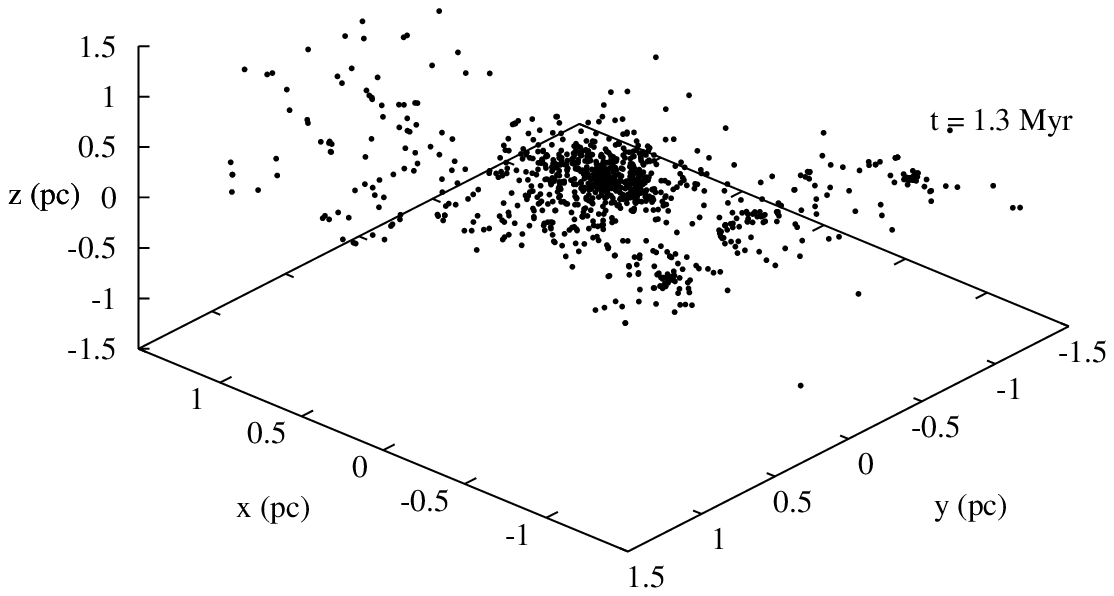}
\\  \includegraphics[height=7.2cm,width=9cm]{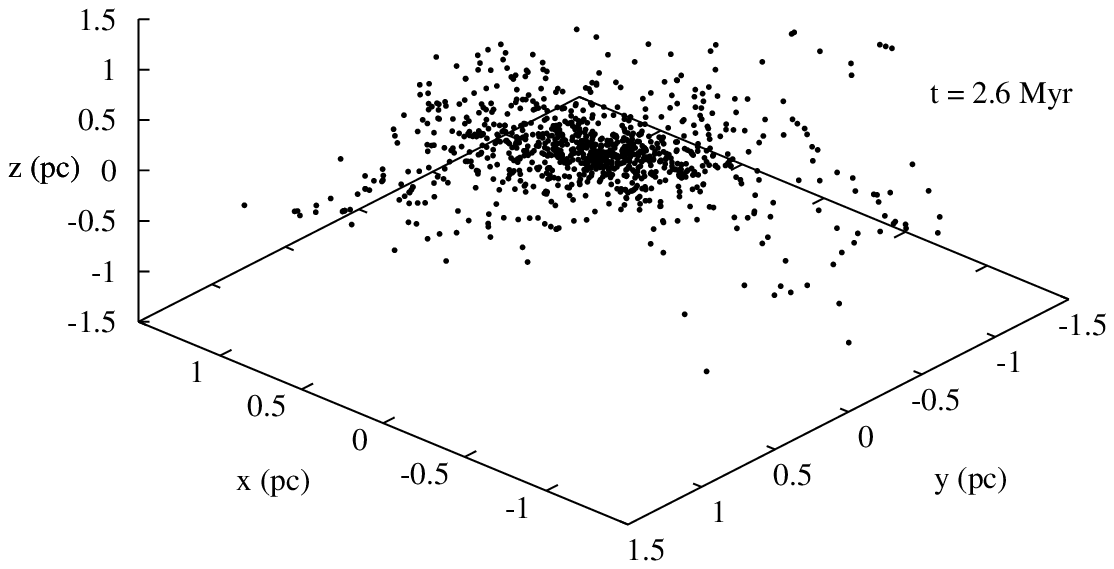} &
\includegraphics[height=7.2cm,width=9cm]{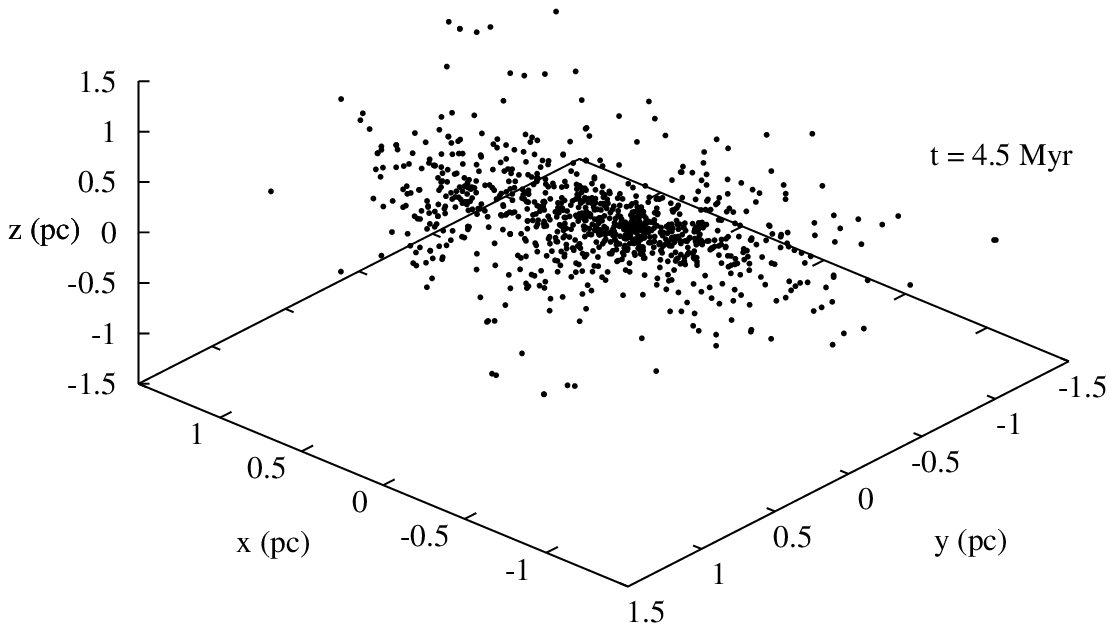}
\\ \includegraphics[height=7.2cm,width=9cm]{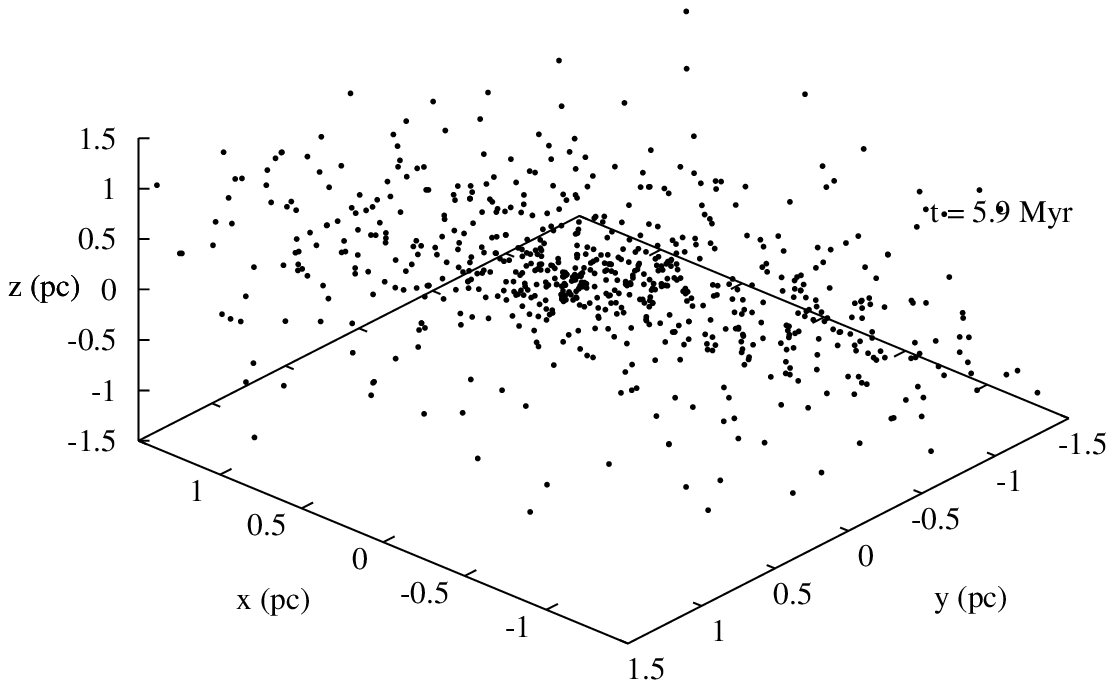} &
\includegraphics[height=7.2cm,width=9cm]{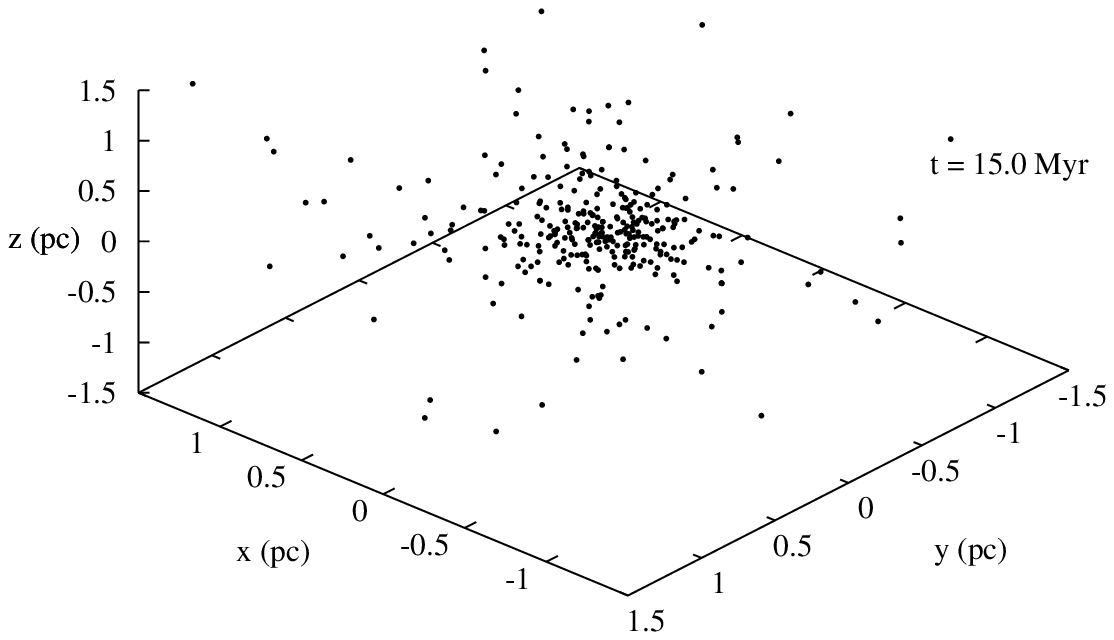} \\
\end{array}$
\end{center}
\caption{The evolution of an initially fractal ($D=1.6$) stellar
  distribution in an $r_{\rm{g}} = 1$~pc, $M_{\rm{g}} = 3450 M_\odot$
  gas potential.  The initial virial ratio of the stellar distribution
  within the gas background potential is $Q_{\rm{i}}$=0.2. The time of
  each snapshot is indicated in the top-right hand corner of each
  panel. Initial clumpy substructure is erased with
  time. Instantaneous gas expulsion occurs at 5~Myr, resulting in
  unbound stars. By 15~Myr the final cluster is settled.}
\label{reprun}
\end{figure*}

The \cite{Baumgardt2007} form is shown on Figure \ref{gasmasslossfig} as
red curves (labelled BK07). We see that our slowest gas mass reduction
($t_{\rm{dur}}$=50~Myr) can be considered approximately equivalent to
the gas mass loss in \cite{Baumgardt2007} when $\tau_{\rm{m}}$=1-2
crossing-times. We emphasise that while a 50~Myr duration
  (i.e. until 99 per cent gas loss) for gas
expulsion may sound unrealistically lengthy, this is merely a result
of the analytical form that we have used to model gas expulsion.

\subsection{Cluster properties at time of gas expulsion}

For a star-gas cluster which is initially smooth and in 
virial equilibrium (i.e. in dynamical equilibrium) and is no longer
forming stars, its response to
gas expulsion can be characterised by two parameters.  Firstly, the
true SFE, the relative gas mass to stellar mass (which does not
change as the cluster is in equilibrium).  Secondly, the timescale of
gas expulsion from  instantaneous (in less than a crossing time), to
adiabatic (in several crossing times).  See
e.g. \cite{Baumgardt2007}.

However, \cite{Goodwin2009} points out that the important variables
are actually the virial state of the stars at the onset of gas
expulsion and the timescale of gas expulsion.  In an initially
virialised cluster the virial state of the stars can be found directly
from the true SFE. However if the cluster is not in equilibrium
initially, then the virial state of the stars corresponds to an {\em
  effective} SFE which can be very different from the true SFE (see
also \citealp{Verschueren1989}; \citealp{Goodwin2006}).

In Smith et al. (2011a), we parameterise the effective SFE using two more
physically meaningful quantities.  The Local Stellar Fraction (LSF) is
a measure of the mass of gas in the region where the stars are found
(a form of evolving effective SFE).  We also define the virial ratio
{\em of the stars} at the moment of gas expulsion $Q_{{\rm ge}}$.

The LSF is defined as
\begin{equation}
\label{LSFeqn}
{\rm LSF}=\frac{M_\star(r<r_{\rm{h(s)}})}{M_\star(r<r_{\rm{h(s)}}) +
  M_{\rm{g}}(r<r_{\rm{h(s)}})}
\end{equation}
where $r_{\rm{h(s)}}$ is the half-mass radius of the stars, and
$M_\star$ and $M_{\rm{g}}$ are the mass of stars and gas,
respectively.  

The key aspects of the LSF and $Q_{{\rm ge}}$ are that they evolve
with the dynamical evolution of the cluster.  In Smith et al. (2011a) we
found the LSF and $Q_{{\rm ge}}$ (measured when gas expulsion begins) to be effective predictors of the
response of an initially out-of-equilibrium cluster to {\em
  instantaneous} gas expulsion after a few crossing times (in their
case 2.5 initial crossing times).

We measure the survival of clusters to gas expulsion by measuring the fraction of stars that are bound to the cluster at $t=15$~Myr. Where sub-structure remains following gas expulsion, we choose the most massive sub-clump to be the main cluster. We measure the number of stars that are bound to the cluster within the inertial frame of the cluster. This is because some clusters have a net velocity though simulation space following gas expulsion. To calculate the average velocity, we use only stars within a specific radius of the cluster centre. We vary the radius to ensure our final bound fraction is not sensitive to the radius we choose. When measuring the bound fraction we only include stars that are bound to the main cluster, and exclude stars that are bound to small subclumps in any remaining substructure.

\subsection{Summary}

We set-up clumpy $500 M_\odot$ star clusters with $N=1000$ equal-mass
particles within a static background gas potential.  The initial
dynamical state of the stars varies from cool and subvirial
($Q_{\rm{i}}$=0.2) to virialised ($Q_{\rm{i}}$=0.5). We then allow the
stars to  dynamically evolve within the gas potential.  Gas expulsion
begins at a time $t=t_{\rm{exp}}$. We vary $t_{\rm{exp}}$ from 0.1~Myr
(essentially immediate gas expulsion) to 9~Myr ($\sim$ 7 crossing
times). At the {\em start} of gas expulsion, we
measure the Local Stellar Fraction of the cluster (LSF), and stellar
virial ratio ($Q_{\rm{ge}}$). We also vary the rate at which gas expulsion occurs by setting the parameter $t_{\rm{dur}}$ (the amount of time required for $99\%$ of the gas to be expelled). We consider 4 values: $t_{\rm{dur}}$ = 0.01, 1.0, 5.0, \& 50~Myr. We measure the final bound fraction
of each star cluster at $t=15$~Myr to determine  how well the cluster
survives. 

\section{Results}

\begin{figure}
  \centering \epsfxsize=8.5cm \epsfysize=6cm
  \epsffile{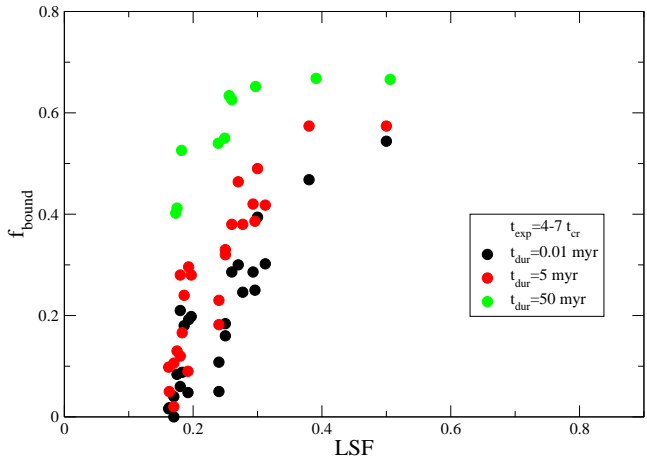}
  \caption{Final bound fraction of stars $f_{\rm{bound}}$ versus Local
    Stellar Fraction (LSF) for clusters where $t_{\rm{exp}}$=4-7~$t_{\rm{cr}}$. Symbol colour denotes the duration over which gas
    expulsion occurs (see legend). A trend for increasing
    $f_{\rm{bound}}$ with increasing LSF can be seen. If gas expulsion
    occurs more slowly, this trend shifts vertically upwards on the
    diagram. By eye, our slowest gas removal rate appears to result in
    an approximate increase in $f_{\rm{bound}}$ of $\sim0.3$.}
\label{tdur}
\end{figure}

\subsection{A representative example}
We will first present a representative example, before focussing on
the impact of the gas expulsion times scales. In Figure \ref{reprun}
we present an example  of the evolution of an initially fractal
$D=1.6$ stellar distribution with an initial virial ratio of $Q_{\rm
  i}=0.2$ in the gas potential. The true SFE of this system is
$\epsilon=0.2$. In this particular example gas expulsion begins at
5~Myr, and then the gas is removed effectively instantaneously.  The
crossing-time of the region is $\sim 1.3~$Myr. 

Figure~\ref{reprun} shows that the stellar component collapses and
relaxes during the first two crossing times.  The initial fractal
distribution (first panel) is partly erased by one crossing time
(second panel) and by two crossing times (third panel) the bulk of the
cluster has collapsed into a dense core (cf. \citealp{Allison2009b};
 Smith et al. 2011a,b).  At the onset of gas
expulsion at  4.5~Myr (fourth panel) the stellar distribution is
relatively relaxed with a halo of stars, mostly ejected by two-body
interactions within the dense main stellar cluster.  The effect of gas
expulsion is as expected in that a significant fraction of stars are
lost.  At 5.9~Myr (fifth panel) many of these escaping stars are still
associated with the cluster, but by 15~Myr (last panel) only a
virialised  bound core of stars remains.  

The cool, clumpy initial distribution of the stars in this simulation
cause them to collapse towards the centre of the gas potential.  This
causes the stars to virialise (i.e. $Q_{\rm ge}$ to approach $0.5$),
but it also causes the LSF to increase significantly as the stars are
concentrated in the centre of the gas potential. For example, in
Figure~\ref{reprun} the initial distribution of stars has an LSF=0.24
(close to the true SFE), but the final LSF at the time of gas
expulsion has increased to 0.41.

\subsection{Effects of varying the rate of gas expulsion}
\label{sectionrateofgasexp}

In all simulations, we evolve our stellar initial conditions  in the
gas potential until gas expulsion begins. At the start of gas
expulsion we measure the LSF (see Equation \ref{LSFeqn}) and the
virial ratio $Q_{\rm{ge}}$. 

\begin{figure}
  \centering \epsfxsize=8.5cm \epsfysize=5.1cm
  \epsffile{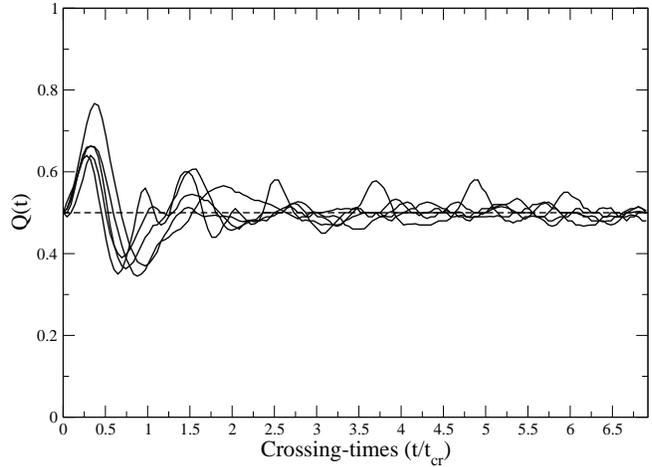}
  \caption{Evolution of the stellar virial ratio of 5 different
    fractal clusters. All start with a total virial ratio of $Q_{\rm
      i} = 0.5$.}
\label{virialwobbles}
\end{figure}

We simulate the evolution of 10 different realisations. For each ensemble
we model gas expulsion occurring at $t_{\rm{exp}}$= 0.1, 1.0, 5.0 \&
9.0 Myr. For each chosen $t_{\rm{exp}}$, we model four durations over
which the gas is expelled: $t_{\rm{dur}}$= 0.01, 1.0, 5.0, \&
50~Myr. We use $Q_{\rm{i}}$=0.2 and 0.5. In total, we conduct 90
simulations. All simulations are conducted for 15 Myr in total ($\sim
11.5$ {\em initial} crossing-times), when the final bound fraction
$f_{\rm{bound}}$ of the cluster is measured. 

The results of the $t_{\rm{exp}}$=5.0-9.0~Myr
($t_{\rm{exp}}$=4-7~crossing-times) simulations are shown on
Figure \ref{tdur} (we turn later to the effects of different 
$t_{\rm{exp}}$).   We plot the final bound fraction $f_{\rm{bound}}$
of the cluster against the LSF at the start of gas expulsion.  The
colour of the filled circle symbols denotes the duration over which
their gas was removed.  For clarity in Figure \ref{tdur} , we exclude the
$t_{\rm{dur}}=1$~Myr points, this gas removal timescale is effectively
  instantaneous and the results are indistinguishable from the 
$t_{\rm{dur}}=0.01$~Myr points. Black symbols have almost instantaneous
gas expulsion ($t_{\rm{dur}}$=0.01~Myr). Red symbols lose their gas
marginally more slowly ($t_{\rm{dur}}$=5~Myr), and green symbols lose
their gas the most slowly of all ($t_{\rm{dur}}$=50~Myr).  (For
reference see Fig. \ref{gasmasslossfig}.)

A trend can be seen in the relation between $f_{\rm{bound}}$ and  the
LSF in Figure \ref{tdur}.  Firstly, as expected, lower LSFs tend to
result in lower bound fractions.  Also, as expected, slower gas
removal (tending to adiabatic) tends to result in higher bound fractions
(cf. \citealp{Baumgardt2007}).  

However, for low-LSFs there is a wide
spread in the final bound fractions.  For LSFs of $\sim 0.2$ (similar 
to the initial, true, SFE), the bound fractions vary between zero and
over half.  The largest bound fractions (all $>0.4$) occur for
adiabatic gas loss which is far less destructive (see above), but for
an intermediate gas loss timescale of $5$~Myr the spread is between
zero and 0.3.  We will return to this in the next section.

We do find broad agreement with the results of \cite{Baumgardt2007}
once the differences in our parameterisation of the gas expulsion
timescale and our use of the LSF rather than true SFE are taken into
account.  Our LSF values are typically between 0.2 and 0.4. We
therefore compare with the \cite{Baumgardt2007} results for true SFEs
in the same range.  The upper-left panel of Figure 2 of
\cite{Baumgardt2007} shows that their bound fraction also increases by
20-40 per cent between their effectively instantaneous and their
adiabatic gas expulsion timescales.  Inspection of Figure \ref{tdur}
shows a similar relationship, albeit with {\em much} more scatter.  

\subsection{The evolution of the dynamical state of the stars}
\label{section_dynamical_evolution}

The scatter we find in the LSF-$f_{\rm{bound}}$ relationship
for a given gas expulsion onset and duration comes from 
two sources.  It is due to differences
in the dynamical state of the stars at the onset of gas expulsion
which, in turn, is due to the stochastic evolution of the initially
non-equilibrium stellar distribution.

Figure~\ref{virialwobbles} shows the evolution of the {\em stellar}
virial ratios for 5 different fractals that initially have a total
virial ratio of $Q_{\rm i} = 0.5$.  The stellar virial ratio is
the virial ratio of the stars alone. We note that the potential of any individual star has a component from surrounding stars and another component from the surrounding gas potential. Both components are included in the virial ratio calculation.

Although the systems are initially
in virial equilibrium,  the fractal initial conditions mean that they
are {\em not} in dynamical  equilibrium.  The stellar distributions
violently relax and collapse which initially increases the stellar
virial ratio before finding a roughly smooth, dense new configuration
in the centre of the gas potential (with a correspondingly higher LSF).
The stars then `bounce', oscillating around virial equilibrium for several
crossing times whilst they relax.

Smith et al. (2011a) found that the stellar virial ratio at the onset of
gas expulsion is important to the cluster's response to gas expulsion
(see also \citealp{Goodwin2009}).  If the stellar virial ratio is
slightly super-virial ($Q_{\rm ge} > 0.5$), then the cluster will
loose more stars (lower $f_{\rm{bound}}$), whilst if they are slightly
sub-virial ($Q_{\rm ge} < 0.5$), then the cluster will retain more
stars (higher $f_{\rm{bound}}$).  However, Smith et al. (2011a) only
examined instantaneous gas expulsion after $2.5$ crossing times.

\begin{figure}
  \centering \epsfxsize=8.5cm \epsfysize=5cm \epsffile{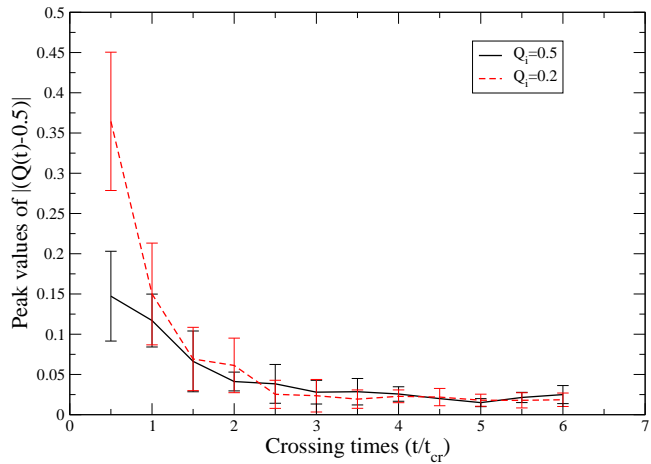}
  \caption{Evolution of peak values of $\lvert ({\rm{Q(t)}}-0.5)
    \rvert$ for 10 different initially virialised ($Q_{\rm{i}}$=0.5)
    fractals and 10 different cool ($Q_{\rm{i}}$=0.2) fractals. The
    average (and standard deviation) of the peak values are binned
    into half crossing-time time-bins. Peak values are initially
    larger, then reduce in size as the clusters attempt to
    virialise. Initially $Q_{\rm{i}}$=0.2 fractals have larger
    deviations from virialised. After $\sim~1.5$ crossing-times,
    $Q$-evolution of $Q_{\rm{i}}$=0.2 $\&$ 0.5 are similar. After 4-5
    crossing-times, fractals have settled to close to virialised.}
\label{qpeaks}
\end{figure}

To investigate the effects of $Q_{\rm ge}$ with different starting
times and durations of gas expulsion we record the peak values of the
deviation of the stellar distribution from virial equilibrium ($
\lvert (Q-0.5) \rvert $) and the time at which these maximum
deviations  occur.   We conduct this test both for 10 different
clusters that are initially globally virialised ($Q_{\rm{i}}$=0.5),
and 10 different clusters that are initially globally cool
($Q_{\rm{i}}$=0.2).  We  calculate the mean and standard deviation of
the peak values of  $ \lvert (Q-0.5) \rvert $ in bins of half
an initial crossing-time. The results are presented in Figure
\ref{qpeaks}.

\begin{figure*}
\begin{center}$
\begin{array}{c}
\includegraphics[height=6.4cm]{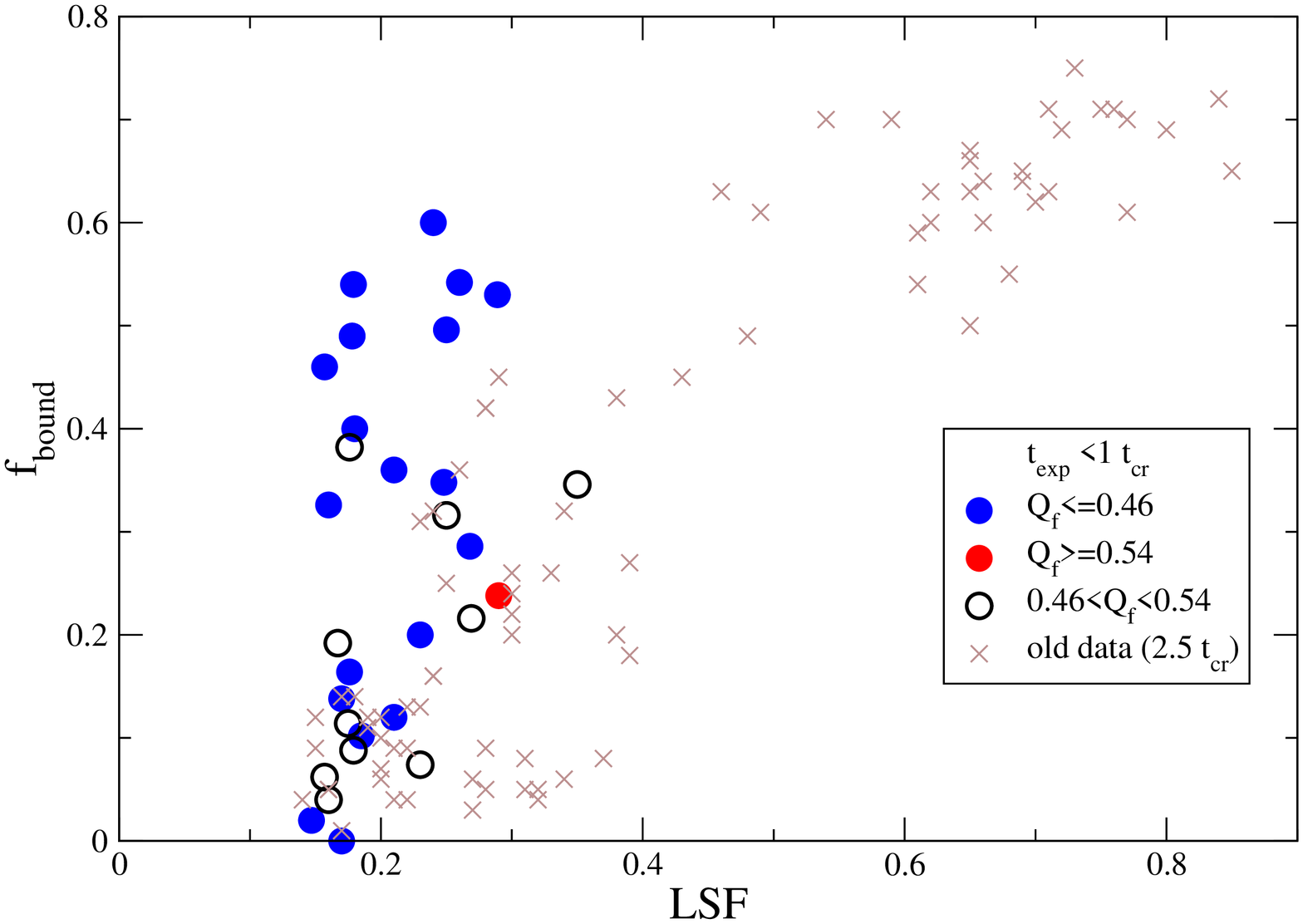} \\
\\ \\
\includegraphics[height=6.4cm]{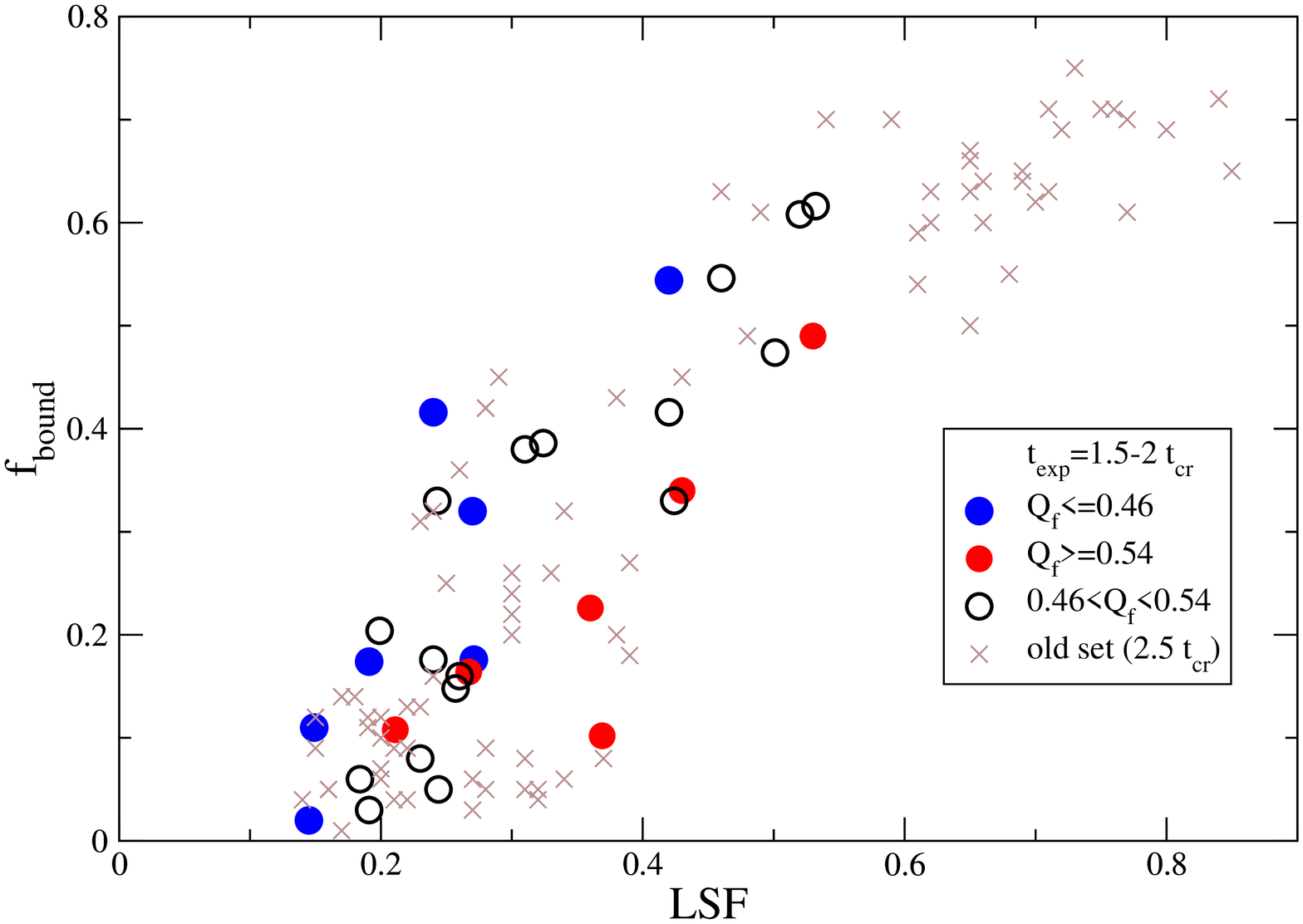} \\
\\ \\
\includegraphics[height=6.4cm]{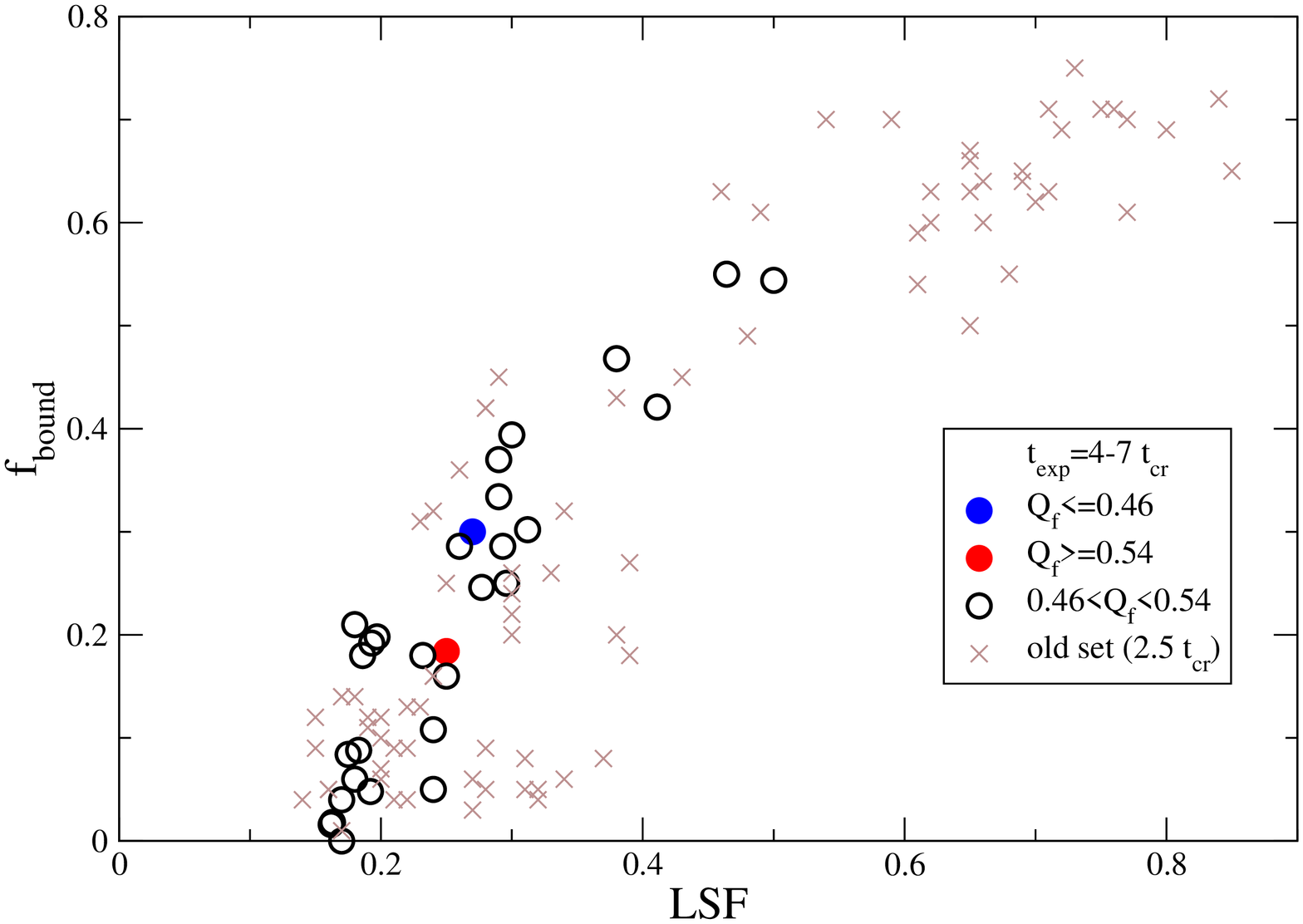} \\
\end{array}$
\end{center}
\caption{Final bound fraction of stars $f_{\rm{bound}}$ versus Local
  Stellar Fraction (LSF) for clusters with early gas removal
  ($t_{\rm{exp}}< 1~t_{\rm{cr}}$, upper panel), intermediate gas
  removal time ($t_{\rm{exp}}$=1.5-2.0~$t_{\rm{cr}}$, middle panel),
  and late gas removal ($t_{\rm{exp}}$=4-7~$t_{\rm{cr}}$, lower
  panel). In these models, when gas expulsion begins, it is
  effectively instantaneous ($t_{\rm{dur}}$=0.01~Myr). Clusters with
  $Q_{\rm i}$=0.5 result in lower LSF values, whereas $Q_{\rm i}$=0.2
  clusters result in higher LSF values. The stellar virial ratio at
  the onset of gas expulsion $Q_{\rm ge}$  is either low (blue
  circles, $Q_{\rm ge} < 0.46$), roughly virialised (open circles,
  $0.46 < Q_{\rm ge} < 0.54$), or high (red squares, $Q_{\rm ge} >
  0.54$).  Light crosses show the results from Smith et al. (2011a) for
  instantaneous gas expulsion at $t_{\rm{exp}}=2.5~t_{\rm{cr}}$ for
  reference. }
\label{fbvslsf_texp}
\end{figure*}

Figure \ref{qpeaks} quantifies what is visible in
Figure~\ref{virialwobbles}.  Deviations from virialised are greater at
early times as the stellar component relaxes.  They are also greater
in initially globally cool clusters ($Q_{\rm{i}}$=0.2) as they have to
relax more to reach (rough) equilibrium.

We can predict that the stellar virial ratio will be an important factor in that it will vary more, and be more stochastic, at early times.  Firstly, the further
clusters start away from equilibrium (dynamical or virial), the greater
the initial variations can be.  Secondly, the shorter the amount of time
clusters have had to relax, the greater the variations can be.

\begin{figure}
  \centering \epsfxsize=8.5cm \epsfysize=5cm \epsffile{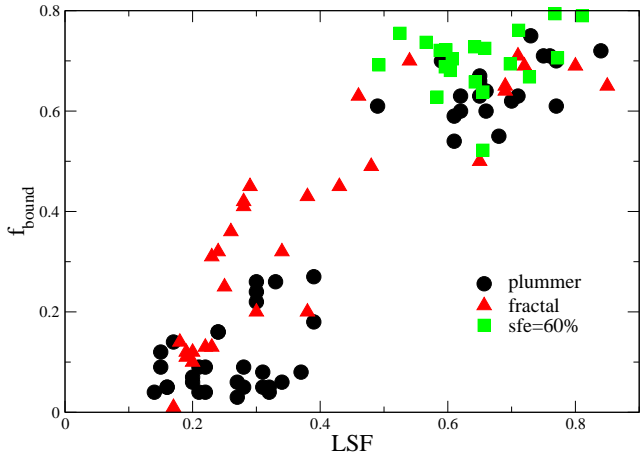}
  \caption{Final bound fraction of stars $f_{\rm{bound}}$ versus Local
  Stellar Fraction (LSF) for clusters with instantaneous gas expulsion at $t_{\rm{exp}}=2.5~t_{\rm{cr}}$. Results from Smith et al. (2011a) where the star formation efficiency is 20$\%$, and using an initial distribution of stars that is `Plummer-like' (black circles), and fractal (red triangles). New results are shown for clusters with matching properties to the fractals (red triangles), but a much higher star formation efficiency of 60$\%$ (green squares). The scatter in the bound fraction is not highly sensitive to the form of the initial clumpy stellar morphology we choose, nor to the star formation efficiency.}
\label{sfedependency}
\end{figure}

\subsection{The survival of clusters}
\label{section_fbound_LSF}

From the above results and discussions we can see that the response of
a cluster to gas expulsion, and the possible survival of a bound core
as a remnant cluster, depend on several factors.

The relative importance of the gas potential to the stars (as measured
by the LSF) is clearly very important.  This, however, depends on the
way in which the stellar distribution has relaxed relative to the
  gas and depends on the initial distribution (in this case highly
clumpy) and initial velocity dispersion (measured by the total initial
virial ratio $Q_{\rm i}$) of the stars.  It also depends on the exact
dynamical state of the stars at the moment when gas expulsion occurs
($Q_{\rm{ge}}$).  And both the LSF and $Q_{\rm{ge}}$ depends to some
extent on the stochastic details of the relaxation of each particular
realisation (see \citealp{Allison2010}).  The reaction of the
cluster to gas expulsion also depends on the timescale of gas
expulsion, with long (adiabatic) timescales being much less
destructive.

In addition, as we will explore in this section, both the LSF and
especially $Q_{\rm{ge}}$ evolve rapidly,  especially in the first 1 --
2 crossing-times and so the effect of gas expulsion will depend
strongly on the time of the onset of gas expulsion.

We have already investigated the effect of varying the rate of gas
expulsion in Section~\ref{sectionrateofgasexp}. We now focus on the
effect of varying the instant at which gas expulsion occurs, and fix
the rate of gas expulsion to be effectively instantaneous.

Figure \ref{fbvslsf_texp} shows the LSF-$f_{\rm{bound}}$ relationships
for early gas expulsion (top panel, $t_{\rm{exp}}<1~t_{\rm{cr}}$),
intermediate gas expulsion (middle panel, $t_{\rm{exp}}=1.5-2.0~t_{\rm{cr}}$), and late gas expulsion (bottom panel,
$t_{\rm{exp}}=4-7~t_{\rm{cr}}$). For all data points shown, when gas
expulsion begins, the gas is removed instantaneously. Plotted on each
panel in light crosses are the results from Smith et al. (2011a)
instantaneous gas expulsion at $t_{\rm{exp}}=2.5~t_{\rm{cr}}$ for
reference.  We combine the results for $Q_{\rm i}$=0.5 and
$Q_{\rm i}$=0.2 star clusters in Figure \ref{fbvslsf_texp}.  

The points marked with blue filled circles have a low stellar virial
ratio at the moment of the onset of gas expulsion ($Q_{\rm ge} <
0.46$), open circles are roughly virialised ($0.46 < Q_{\rm ge} <
0.54$), and red squares are super-virial ($Q_{\rm ge} > 0.54$).

The top panel shows the results for the early onset of gas expulsion.
Early gas expulsion results in gas expulsion occurring whilst the
stars are still relaxing.  They tend to have a low virial ratio
($Q_{\rm ge} < 0.46$) as they are still collapsing into a new
equilibrium, and their LSF tends to be low (close to the true SFE) as
they are still fairly widely distributed in the gas potential.  The
huge scatter in this top panel is due to the highly stochastic nature
of the early dynamics of the clusters.  Many of these clusters will
still contain subclumps (see Figure~\ref{reprun}) and the bulk
dynamics of these subclumps is important. At these stages
$Q_{\rm ge}$ is not a good indicator of survival as much of the
stellar kinetic energy can be contained in bound clumps.  If these
clumps escape due to gas expulsion $f_{\rm bound}$ will be low (or
even zero), however retaining one or two subclumps can result in a
very high $f_{\rm bound}$. Clumps also raise the stellar-to-gas fraction very locally at their centres. The LSF is insensitive to this as it measures the stellar-to-gas fraction on scales that are typically much larger than individual clumps (i.e. the half-mass radius of the stars).

For intermediate gas expulsion timescales (middle panel) the LSF
becomes more important as the clusters tend to have had the time
to relax into a smooth central cluster.  More clusters are roughly virialised and there is
a trend of $f_{\rm bound}$ decreasing with lower LSF as would be
expected.  As clusters are oscillating around virial equilibrium
within the gas potential, the spread in $f_{\rm bound}$ for a given
LSF is wide. At a fixed LSF, sub-virial clusters tend to higher
$f_{\rm bound}$ than super-virial clusters. Therefore at intermediate
gas expulsion timescales, a combination of $Q_{\rm ge}$ and LSF
determine the response to gas expulsion.

For long gas expulsion timescales (bottom panel) the clusters have had
a chance to virialise within the gas potential and so the relationship
between the LSF and $f_{\rm bound}$ becomes much tighter with the LSF
being the key factor in determining the response to gas expulsion.

\subsection{Dependency on initial stellar morphology and star formation efficiency}

First, we wish to test how sensitive our results are to the mathematical form of the initial clumpy distribution of stars we have chosen. To do this, we use models from Smith et al. (2011a) where we considered two types of clumpy distributions (or stellar morphologies) for the initial positions of our stars; fractal (like those modelled in this paper), and `Plummer-like'. In the `Plummer-like' distribution there are 16 clumps of stars in total, and each clump is modelled as an individual Plummer sphere of mass $\sim30~M_{\odot}$, and Plummer scale-length $\sim0.05$~pc. The clumps themselves are distributed according to a Plummer distribution with a Plummer scale-length of 1-1.5~pc). Further details can be found in Smith et al. (2011a).

In Figure \ref{sfedependency} we show the final bound fraction of stars $f_{\rm{bound}}$ versus Local Stellar Fraction (LSF) for clusters from Smith et al. (2011a), but split into `Plummer-like' (black circles), and fractal (red triangles). We recall that both groups have equal star formation efficiency of $20\%$, and both undergo instantaneous gas expulsion after 2.5 crossing-times - they differ only in their initial stellar morphology. We find that the trend, and its scatter in bound fraction, is not highly sensitive to which form of clumpy stellar morphology we choose. Two apparently separated groups of Plummer-like models are seen at high and low LSF. However this is mainly just a result of low numbers of Plummer-like models with $Q_{\rm{i}}=$0.2-0.4, whose presence would fill the gap between the two groups.

Secondly, we wish to test if our results differ when we consider clusters whose mass is dominated by stars, as we have so far only considered clusters that are heavily dominated by gas. Therefore we model 20 clusters with a much higher star formation efficiency of 60$\%$ (green squares). These differ from the fractals on Figure \ref{sfedependency} in their star formation efficiency only. As expected, these clusters fall in the high LSF range. More importantly, we find that there is still a similar amount of scatter at higher star formation efficiency - despite the fact that the stars now dominate the mass of the cluster. In fact, the virial state at the time of gas expulsion $Q_{\rm ge}$ continues to play the same role, broadening the trend, as when the star formation efficiency was 20$\%$. 

In summary, the trend and its scatter is not very sensitive to the form of the initial clumpy stellar morphology we choose, nor to the star formation efficiency (i.e. the strength of the rigid background potential).

\section{Discussion}

We have examined the response of young clusters to gas expulsion as
measured by their ability to retain a bound core of stars and avoid
complete destruction: their survivability.

Clusters older than $\sim 5$~Myr are not seen to be associated with
the gas from which they formed.  There are two possibilities for the
fate of this gas.  Firstly, that all of the gas was used in star
formation (100 per cent efficiency).  However this is unlikely as shortly after a massive (O- or B-star) forms it will heat the gas and remove
it from the site of star formation (through winds, ionising radiation,
or supernovae).  The removal of a (very significant) fraction of the
(gas) mass in a region will have a (very significant) effect on the
potential felt by the stars that remain.  If the stars are unbound or
very loosely bound at formation they will be dispersed, and gas
expulsion merely speeds this process somewhat.  However, if the stars
are bound then gas expulsion can completely destroy the cluster, or
leave a bound core which will be seen as a naked cluster.  It must
happen that bound cores are left after gas expulsion as many old,
bound clusters are seen (most strikingly, globular clusters).

A key question to address is what can be said about the
initial conditions of old, bound clusters?  What must be true is 
that they have survived gas expulsion and therefore they must have 
been bound at the onset of gas expulsion.  

Most previous work on gas expulsion has concentrated on clusters that
are in virial and dynamical equilibrium at the onset of gas expulsion
(see e.g. \citealp{Tutukov1978}; \citealp{Hills1980}; \citealp{Goodwin2006}; \citealp{Baumgardt2007}).  In such a situation the primary factor controlling the
survival of a cluster is the true SFE, secondary factors are the
timescale of gas expulsion (adiabatic gas expulsion is less
destructive), and the tidal radius (expansion over the tidal radius
can rapidly destroy a cluster), as shown in detail by \cite{Baumgardt2007}.

However, observations of star formation lead us to believe that stars
do not form in equilibrium.  They certainly seem to form in a complex
distribution that follows the gas (e.g. \citealp{Larson1995}; \citealp{Testi2000}; \citealp{Cartwright2004}; \citealp{Gutermuth2005}; 
\citealp{Allen2007}; \citealp{Kirk2007};
\citealp{Schmeja2008}; \citealp{Gutermuth2009}; \citealp{Peretto2009}; \citealp{Bressert2010}; \citealp{Difrancesco2010}; \citealp{Maury2011}), they seem to form with a lower velocity dispersion than
the gas (\citealp{Andre2002}; \citealp{Walsh2004}; \citealp{Adams2006}; 
\citealp{Peretto2006}; \citealp{Kirk2007}; \citealp{Proszkow2009}), and 
their early evolution appears to be dramatic
(e.g. \citealp{Bastian2008}; \citealp{Allison2009b}; \citealp{Allison2010}).

In this paper we have shown that if the initial {\em stellar}
distribution is out-of-equilibrium then the effect of gas expulsion
can be very stochastic (see also Smith et al. 2011a,b).  In particular,
there are several key factors:\\
1. The effective SFE (measured by us as the LSF): the relative
  stellar and gas masses at the onset of gas expulsion (see also
  \citealp{Verschueren1989}; \citealp{Goodwin2009}).  The higher the effective
  SFE, the greater the survivability.\\
2. The stellar virial ratio ($Q_{\rm ge}$): the virial state of the
  {\em stars} alone at the onset of gas expulsion.  The lower the
  stellar virial ratio, the greater the survivability.\\
3. The time of the onset of gas expulsion: this effects to what
extent the stars and gas can relax and/or decouple and so changes the
effective SFE and stellar virial ratio.  From our simulations, the
earlier the onset of gas expulsion the greater the stochasticity in
survivability.\\ 
4. The timescale of gas expulsion: instantaneous gas expulsion is more
destructive than adiabatic (see also \citealp{Baumgardt2007}).  \\
5. Tidal radius: whilst we have not simulated this, there is no reason
to think it should not have the same effect as that found by \citealp{Baumgardt2007}.

The first four of these factors are all stochastic and depend on the
details of the initial conditions.

We reiterate here that our initial conditions are not realistic.  We evolve
equal-mass stars in a static background potential which we then remove in a
simplistic way.  However, we find it impossible to see how more
realism could possibly produce {\em less} stochasticity.  Indeed, more
realistic initial conditions should introduce vast quantities of variables that are potentially important.

For example, a realistic clumpy, live gas potential could react in very
different ways.  Depending on the details of the gas it could couple
well to the stars (i.e. lower the LSF by moving with the stars), or
couple badly (by being accreted or removed from subclumps, e.g. 
\citealp{Kruijssen2012}).  Depending on the structure of the gas, feedback could be
less efficient, possibly escaping through bubbles (e.g. \citealp{Dale2011}),
changing the onset of significant gas expulsion and/or its timescale.
In addition, realistic small-$N$ subclumps would contain a mass
spectrum of stars, include binaries, and have a wealth of possible
$N$-body interactions that are extremely stochastic on their own
(\citealp{Allison2010}).

In summary, there is no reason to think that if one cluster survives
and another does not that they had very different initial conditions.
Two sets of {\em statistically} identical initial conditions can produce a cluster that
retains most of its stars or one that is completely destroyed.  The
final outcome depends on a wealth of different variables (only a
handful of which we have rather simplistically modelled here), as well
as the details of the initial conditions.

\section{Summary $\&$ Conclusions}

We perform $N$-body simulations of sub-structured, non-equilibrium
$2500 M_\odot$ clusters with $500 M_\odot$ of stars in $N=1000$ 
equal-mass particles. A static
background potential provides a gas mass of $2000 M_\odot$ surrounding the stars. The initial true 
star formation efficiency is
fixed at 20 per cent.  Gas expulsion is modelled as a reduction in the mass
of the background potential over time. We allow the instant that gas
expulsion begins, and the duration over which gas expulsion occurs, to
vary. 

We use two parameters to characterise the state of the cluster at the
moment of gas expulsion.  The Local Stellar Fraction (LSF) is a
measure of the relative contribution of gas and stars to the potential
felt by the stars.  The stellar virial ratio is a measure of how
relaxed the stellar component is.

Our key results may be summarised as follows.
\begin{enumerate}
\item In initially clumpy, and non-equilibrium clusters, a slower rate
  of gas loss increases the survival of the cluster (in keeping with
  previous results).
\item If gas is expelled within an initial crossing time the effect of
  gas expulsion is completely stochastic and there are no good global 
  indicators of its effects.
\item If gas is expelled after a few crossing times then a combination
  of the LSF (or rather the efficiency of the coupling of the gas to
  the stars), and stellar virial ratio are required to parameterise
  cluster survival, although it is still rather stochastic.
\item If gas is expelled after several crossing times the stellar
  component is generally relaxed and the LSF alone is a good indicator
  of the effects of gas expulsion.  However, the LSF can substantially differ
  from the true star formation efficiency due to the decoupling of the
  gas and stars.
\end{enumerate}

Our simulations have been very simple numerical experiments. However
we feel they do capture some important physics.

Firstly, the coupling of the stars and gas is crucial to the survival
of a cluster.  If the stars are able to relax to a more concentrated
distribution than the gas (or conversly, if the gas is able to expand
relative to the stars), then the dramatic effects of gas expulsion are
reduced.  

Stochasticity is important:  the exact level of decoupling from the gas, and
the details of the stellar phase space distribution are also crucial.
Gas expulsion whilst the stars are still `clumpy' can completely
destroy one cluster, whilst leaving another -- statistically identical
-- cluster with most of its stars in a bound cluster.  This depends on
the bulk virial ratio of the stars, and also the details of the clump
velocities.  It is only after several initial crossing times -- after 
a cluster has relaxed into a rough equilibrium -- that stochasticity
reduces in importance.  However even at this point, stochasticity in the earlier
evolution can still influence the final equilibrium configuration.

We would argue that more realistic initial conditions and physics
(e.g. a mass spectrum of stars, a clumpy live gas potential, feedback,
etc.) can only work to make the problem {\em more} stochastic, rather
than less.  Therefore we contend that it is impossible to give a
`standard' outcome for any statistically identical clusters: some will
be destroyed, while others will survive with different mass and structure.  For this reason it is impossible to go backwards from an
observed surviving cluster to its initial conditions except in a very
broad statistical sense.  Therefore we urge caution in deriving the
initial cluster population from an observed naked cluster population.

\section*{Acknowledgements}
MF was acknowledges support through FONDECYT grant 1095092, RS was financed through a combination of FONDECYT grant 3120135 and a COMITE MIXTO grant, and PA was financed through a CONCICYT PhD Scholarship.

\bibliography{bibfile}

\bsp

\label{lastpage}

\end{document}